\documentclass[%
 reprint,
 amsmath,amssymb,
 aps,
]{revtex4-1}

\usepackage{graphicx}
\usepackage{dcolumn}
\usepackage{bm}

\begin{document}

\title{Universal behavior of extreme value statistics for selected observables of dynamical systems}
\author{Valerio Lucarini}
\email{Email: \texttt{valerio.lucarini@zmaw.de}}
\altaffiliation{Also at: Department of Mathematics and Statistics, University of Reading, Reading, UK.}
\author{Davide Faranda}%
\author{Jeroen Wouters}%
\affiliation{Klimacampus, University of Hamburg, Grindelberg 5, 20144, Hamburg, Germany}%

%


\date{\today}

\begin{abstract}
The main results of the extreme value theory developed for the investigation of the observables of dynamical systems rely, up to now, on the Gnedenko approach. In this framework, extremes are basically identified with the block maxima of the time series of the chosen observable, in the limit of infinitely long blocks. It has been proved that, assuming suitable mixing conditions for the underlying dynamical systems, the extremes of a specific class of observables are distributed according to the so called Generalized Extreme Value (GEV) distribution. Direct calculations show that in the case of quasi-periodic dynamics the block maxima are not distributed according to the GEV distribution. In this paper we show that, in order to obtain a universal behaviour of the extremes, the requirement of a mixing dynamics can be relaxed if the Pareto approach is used, based upon considering the exceedances over a given threshold. Requiring that the invariant measure locally scales with a well defined exponent - the local dimension -, we show that the limiting distribution for the exceedances of the observables previously studied with the Gnedenko approach is a Generalized Pareto distribution where the parameters depends only on the local dimensions and the value of the threshold. This result allows to extend the extreme value theory for dynamical systems to the case of regular motions. We also provide connections with the results obtained with the Gnedenko approach. In order to provide further support to our findings, we present the results of numerical experiments carried out considering the well-known Chirikov standard map.
\end{abstract}

\pacs{Valid PACS appear here}
\maketitle


\section{Introduction}

Extreme value Theory was originally introduced by \citet{fisher} and formalised by \citet{gnedenko}, who showed that the distribution of the maxima of a sample of independent identically distributed (i.i.d) stochastic variables converges under very general conditions to a member of the so-called Generalised Extreme Value (GEV) distribution. The attention of the scientific community to the problem of understanding extreme values theory is growing, also because this theory is crucial in a wide class of applications for defining risk factors such as those related to instabilities in the financial markets and to natural hazards related to seismic, climatic and hydrological extreme events. Even if the probability of extreme events decreases with their magnitude, the damage that they may bring increases rapidly with the magnitude as does the cost of protection against them. From a theoretical point of view, extreme values of observables are related to large fluctuations of the corresponding underlying system. An extensive account of recent results and relevant applications is given in \citep{ghil2010extreme}.

The traditional ('Gnedenko') approach for the statistical inference of extremes is related to the original results by \citet{gnedenko}: we partition the experimental time series into bins of fixed length, we extract the maximum of each bin, and fit the selected data to the GEV distribution family using methods such as maximum likelihood estimation (MLE) or L-moments. See \citep{felici1} for a detailed account of this methodology. The selection of just one maximum in a fixed period may lead to the loss of relevant information on the large fluctuations of the system, especially when there are many large values in a given period \citep{castillo1997fitting}. This problem can be taken care of by considering several of the largest order statistics instead of just the largest one. For such maxima distributions we expect convergence to the  Generalized Pareto Distribution (GPD) introduced by \citet{pickands1975statistical} and \citet{balkema1974residual} to model the exceedances over a given threshold. We call this the 'Pareto' approach. Also this approach has been widely adopted for studying empirically natural extreme phenomena such as those related to waves, winds, temperatures, earthquakes and floods \citep{simiu1996extreme, bayliss1993peaks,pisarenko2003characterization}.

Both the Gnedenko and Pareto approaches were originally designed to study extreme values for series of i.i.d variables. In this case it is well known that a strong connections exists between the two methodologies, as we have that if block maxima obey the GEV distribution, then exceedances over some high threshold will have an associated GPD. Moreover, the shape parameter of the GPD and that of the corresponding GEV distribution are identical \citep{leadbetter}. As a result, several practical methods (e.g. Hill's and Pickands' estimators) developed for estimating the shape parameter of the GEV distribution of the extremes of a given time series are actually based upon comparing the GPD fits at various thresholds \citep{gpd1,coles2001introduction}. In practical terms, it appears that, while the Gnedenko and Pareto approaches provide equivalent information in the asymptotic limit of infinitely long time series, the GPD statistics is more robust when realistic, finite time series are considered (see, e.g., \citep{ding2008newly}).

In recent years, especially under the influence of the rapid development of numerical modelling in the geophysical sciences and of its applications for the investigation of the socio-economic impacts of extreme events, it has become of great relevance to understand whether it is possible to apply the extreme value theory on the time series of observables of deterministic dynamical systems. Carefully devised numerical experiments on climate models of various degrees of complexity have shown that the speed of convergence (if any) of the statistical properties of the extremes definitely depends on the chosen climatic variable of interest \citep{kharin2005,vannitsem,felici1,vitolo}.

Several papers have addressed this issue at a more general level. A first important result is that when a dynamical system has a regular (periodic of quasi-periodic) behaviour, we do not expect, in general, to find convergence to GEV distributions for the extremes of any observable. These results have been presented by \citet{balakrishnan}, and more recently, by \citet{nicolis} and by \citet{haiman}.

A different mathematical approach to extreme value theory in dynamical systems was proposed in the landmark paper by \citet{collet2001statistics}, which has paved the way for the recent results obtained in the last few years \citep{freitas2008,freitas,freitas2010extremal,gupta2009extreme}. The starting point of all of these investigations has been to associate to the stationary stochastic process given by the dynamical system, a new stationary independent sequence which obeys one of the classical three extreme value laws introduced by \citet{gnedenko}. The assumptions which are necessary to observe a GEV distribution in dynamical systems rely on the choice of suitable observables (specific functions of the distance between the orbit and the initial condition, chosen to be on the attractor) and the fulfillment of particular mixing conditions that guarantee the independence of subsequent maxima. Recent studies have shown that the resulting parameters of the GEV distributions can be expressed as simple functions of the local (around the initial condition) dimension of the attractor, and detailed numerical investigations have clarified the conditions under which convergence to the theoretical GEV distributions can be satisfactorily achieved when considering finite time series \citep{faranda2011generalized,faranda2011numerical,faranda2011extreme}.

In this paper, we wish to attempt a unification of these two lines of work by using the Pareto rather than the Gnedenko approach. We choose the same class of observables presented in \citep{freitas2008,freitas,freitas2010extremal,gupta2009extreme,faranda2011generalized,faranda2011numerical,faranda2011extreme} and show that, assuming only that the local measures scales with the local dimension \citep{Bandt2007}, it is possible to obtain by direct integration a GPD for the threshold exceedences when considering a generic orbit of a dynamical systems, without requiring any special mixing properties. The parameters will depend only on the choice of the threshold and, more importantly, on the local dimension. Note that \citet{castillo1997fitting} had already pointed out that in the case of periodic or quasi-periodic motion the Gnedenko approach to the evaluation of the extreme value statistics is inefficient, basically because in the limit of very large blocks, we tend to observe always the same maximum in all bins. To support our analytical results we provide numerical experiments that we carry out considering the classic Chirikov standard map \citep{ScholarC}. This paper is organised as follows. In section 2 we recapitulate the extreme value theory for dynamical systems obtained using the Gnedenko approach. In section 3 we present our general results obtained using the Pareto approach. In section 4 we provide support to our investigation by examining the results of the numerical simulations performed on the standard map. In section 5 we present our final remarks and future scientific perspectives.


\section{\label{sec:level2} Gnedenko Approach: Generalized Extreme Value distributions in dynamical systems}
\citet{gnedenko}  studied the convergence of maxima of i.i.d.
variables $$X_0, X_1, .. X_{m-1}$$ with cumulative distribution function
(cdf) $F(x)=P\{a_m(M_m-b_m) \leq x\}$ where $a_m$ and $b_m$ are normalizing sequences and $M_m=\max\{ X_0,X_1, ..., X_{m-1}\}$. Under general hypothesis on the nature of the parent distribution of data, \citet{gnedenko} showed that the asymptotic distribution of maxima, up to an affine change of variable, belongs to a single family of generalized distribution called GEV distribution whose cdf can be written as:
\begin{equation}
F_{GEV}(x;\mu,\alpha,\kappa)=\textrm{e}^{-t(x)}
\label{GEV}
\end{equation}
where
\begin{equation}
t(x) = \begin{cases}\big(1+\kappa(\tfrac{x-\mu}{\alpha})\big)^{-1/\kappa} & \textrm{if}\ \kappa\neq0 \\ e^{-(x-\mu)/\alpha} & \textrm{if}\ \kappa=0\end{cases}.
\label{GEV1}
\end{equation}
This expression holds for $1+{\kappa}(x-\mu)/\alpha>0 $, using $\mu \in \mathbb{R}$ (location parameter) and $\alpha>0$ (scale parameter) as scaling constants in place of $b_m$, and $a_m$ \citep{pickands}, in particular, in \citet{faranda2011numerical} we have shown that $\mu=b_m$ and $\alpha=1/{a_m}$, where  ${\kappa} \in \mathbb{R}$ is the shape parameter (also called the tail index). When ${\kappa} \to 0$, the distribution corresponds to a Gumbel type (Type 1 distribution).  When the index is positive, it corresponds to a Fr\'echet (Type 2 distribution); when the index is negative, it corresponds to a Weibull (Type 3 distribution).

Let us consider a dynamical system $(\Omega, {\cal B}, \nu, f)$, where $\Omega$ is the invariant set in some manifold, usually $\mathbb{R}^d$, ${\cal B}$ is the Borel $\sigma$-algebra, $f:\Omega\rightarrow \Omega$ is a measurable map and $\nu$ an $f$-invariant Borel measure. In order to adapt the extreme value theory to dynamical systems, following \citep{freitas2008,freitas,freitas2010extremal,gupta2009extreme}, we consider the stationary stochastic process $X_0,X_1,...$  given by:
\begin{equation}
X_m(x)=g(\mbox{dist}(f^m (x), \zeta)) \qquad \forall m \in \mathbb{N}
\label{sss}
\end{equation}
where 'dist' is a distance in the ambient space  $\Omega$, $\zeta$ is a given point and $g$ is an observable function. The partial maximum in the Gnedenko approach is defined as:
\begin{equation}
{M_m}= \max\{ X_0, ... , X_{m-1} \}.
\label{maxi}
\end{equation}
Defining $r=\mbox{dist}(x, \zeta)$, we consider the three classes  of observables $g_i,i=1,2,3$:
\begin{eqnarray}
g_1(r)= -\log(r) \label{g1} \\
g_2(r)=r^{-\beta} \label{g2} \\
g_3(r)=C -r^{\beta} \label{g3}
\end{eqnarray}
where $C$ is a constant and $\beta>0 \in \mathbb{R}$. Using the observable $g_i$  we obtain convergence of the statistics of the block maxima of their time series obtained by evolving the dynamical system to the Type $i$ distribution if one can prove two sufficient conditions called $D_2$ and $D'$, which basically imply a sort of independence of the series of extremes resulting from the mixing of the underlying dynamics \citep{freitas2008}. The conditions cannot be simply related to the usual concepts of strong or weak mixing, but are indeed not obeyed by observables of systems featuring a regular dynamics or power-law decay of correlations.

A connection also exists between the existence of extreme value laws and the statistics of first return and hitting times, which provide information on how fast the point starting from the initial condition $\zeta$ comes back to a neighborhood of $\zeta$, as shown by \citet{freitas} and \citet{freitasNuovo}. In particular, they proved that for dynamical systems possessing an invariant measure $\nu$, the existence of an exponential hitting time statistics on balls around $\nu$-almost any point $\zeta$ implies the existence of extreme value laws for one of the observables of type $g_i, i=1,2,3$ described above. The converse is also true, namely if we have an extreme value law which applies to the observables of type $g_i, i=1,2,3$ achieving a maximum at $\zeta$, then we have exponential hitting time statistics to balls with center $\zeta$. Recently these results have been generalized to local returns around balls centered at periodic points \citep{freitas2010extremal}.

In \citet{faranda2011numerical,faranda2011extreme,faranda2011generalized} we analised both from an analytical and numerical point of view the extreme value distribution in a wide class of low dimensional maps. We divided the time series of length $k$ of the $g_i$ observables into $n$ bins each containing the same number $m$ of observations, and selected the maximum (or the minimum) value in each of them \citep{coles}. We showed that at leading order (the formulas are asymptotically correct for $m,k\rightarrow\infty$), the GEV parameters in mixing maps can be written in terms of $m$ (or equivalently $n$) and the local dimension of the attractor $D$. We have:
\begin{itemize}
  \item $g_1$-type observable:
\begin{equation}
\alpha= \frac{1}{D} \qquad \mu \sim \frac{1}{D}\ln(k/n) \qquad \kappa=0
\label{g1res}
\end{equation}
  \item $g_2$-type observable:
\begin{equation}
\alpha\sim n^{-\frac{\beta}{D}} \qquad \mu \sim n^{-\frac{\beta}{D}} \qquad \kappa=\frac{\beta}{D}
\label{g2res}
\end{equation}
  \item $g_3$-type observable:
\begin{equation}
\alpha\sim n^{\frac{\beta}{D}} \qquad \mu = C \qquad \kappa=-\frac{\beta}{D}
\label{g3res}
\end{equation}
\end{itemize}
Moreover, we clearly showed that other kind of distributions not belonging to the GEV family are observed for quasi-periodic and periodic motions.
\section{\label{sec:level3}The Pareto approach: Generalized Pareto Distributions in Dynamical Systems}
We define an exceedance as $z=X-T$, which measures by how much $X$ exceeds the threshold $T$. As discussed above, under the same conditions under which the block maxima of the i.i.d. stochastic variables $X$ obey the GEV statistics, the exceedances $z$ are asymptotically distributed according to the Generalised Pareto Distribution \citep{leadbetter}:
\begin{equation}
F_{GPD}{(z;\xi,\sigma)} = \begin{cases}
1- \left(1+ \frac{\xi z}{\sigma}\right)^{-1/\xi} & \text{for }\xi \neq 0, \\
1-\exp \left(-\frac{z}{\sigma}\right) & \text{for }\xi = 0,
\end{cases}
\label{GPD}
\end{equation}
where the range of $z$ is $0 \leq z < \infty$ if $\xi \leq 0$ and $ 0 \leq z \leq \sigma/\xi$ if $\xi>0$. We consider the same set up described in the previous section and take into account an observables $g=g(\mbox{dist}(x, \zeta))=g(r)$, such that $g$ achieves a maximum $g_{max}$ for $r=0$ (finite or infinite) and is monotonically decreasing. We study the exceedance above a threshold $T$ defined as $T=g(r^*)$. We obtain an exceedence every time the distance between the orbit of the dynamical system and $\zeta$ is smaller than $r^*$. Therefore, we define the exceedances $z= g(r)-T$. By the Bayes' theorem, we have that $P(r<g^{-1}(z+T)|r<g^{-1}(T))=P(r<g^{-1}(z+T))/P(r<g^{-1}(T))$. In terms of invariant measure of the system, we have that the probability $H_{g,T}(z)$ of observing an exceedance of at least $z$ given that an exceedence occurs is given by:
\begin{equation}
H_{g,T}(z)\equiv\frac{\nu( B_{g^{-1}(z+T)}(\zeta))}{\nu( B_{g^{-1}(T)}(\zeta))}.
\label{measureball}
\end{equation}
Obviously, the value of the previous expression is 1 if $z=0$. In agreement with the conditions given on $g$, the expression contained in Eq. \eqref{measureball} monotonically decreases with $z$ and vanishes when the radius is given by $g^{-1}(g_{max})$. Note that the corresponding cdf is given by $F_{g,T}(z)=1-H_{g,T}(z)$. In order to address the problem of extremes, we have to consider small radii. At this regard we will invoke, and assume, the existence of the following limit
\begin{equation}
\lim_{r\rightarrow 0}\frac{\log \nu( B_{r}(\zeta))}{\log r}=D(\zeta), \ \mbox{for} \  \zeta \ \mbox{chosen} \ \nu-\mbox{a.e.},
\label{limite}
\end{equation}
where $D(\zeta)$ is the local dimension of the attractor \citep{Bandt2007}.
Therefore, we rewrite the following expression for the tail probability of exceedance:
\begin{equation}
H_{g,T}(z)\sim \left(\frac{g^{-1}(z+T)}{g^{-1}(T)}\right)^D.
\label{measureball1}
\end{equation}
where we have dropped the $\zeta$ dependence of $D$ to simplify the notation. By substituting $g$ with specific observable we are considering, we obtain explicitly the corresponding extreme value distribution law.

By choosing an observable of the form given by either $g_1$, $g_2$, or $g_3$, we derive as extreme value distribution law one member of the Generalised Pareto Distribution family given in Eq. \eqref{GPD}. Results are detailed below:
\begin{itemize}
  \item $g_1$-type observable:
\begin{equation}
\sigma= \frac{1}{D} \qquad \xi=0;
\label{gpd1res}
\end{equation}
  \item $g_2$-type observable:
\begin{equation}
\sigma=\frac{T\beta}{D} \qquad \xi=\frac{\beta}{D};
\label{gpd2res}
\end{equation}
  \item $g_3$-type observable:
\begin{equation}
\sigma=\frac{(C-T)\beta}{D} \qquad \xi=-\frac{\beta}{D}.
\label{gpd3res}
\end{equation}
\end{itemize}
The previous expressions show that there is a simple algebraic link between the parameters of the GPD and the local dimension of the attractor around the point $\zeta$. This implies that the statistics of extremes provides us with a new algorithmic tool for estimating the local fine structure of the attractor. These results show that it is possible to derive general properties for the extreme values of the observables $g_1$, $g_2$, or $g_3$ independently on the qualitative properties of the underlying dynamics, be the system periodic, quasi-periodic, or chaotic. Therefore, by taking the Pareto instead of the Gnedenko approach, we are able to overcome the mixing conditions (or the requirements on the properties of the hitting time statistics) needed to derive a general extreme value theory for dynamical systems, as proposed in \citep{collet2001statistics,freitas2008,freitas,freitas2010extremal,gupta2009extreme}. In \citep{faranda2011numerical,faranda2011extreme,faranda2011generalized} we had proposed that the reason why a link between the extreme value theory and the local properties of the invariant measure in the vicinity of the point $\zeta$ can be explained by the fact that selecting the extremes of the observables $g_1$, $g_2$, or $g_3$ amounts to performing a zoom around $\zeta$. In the case of the Gnedenko approach, such a picture is accurate only if the dynamics is mixing (time and spatial selection criteria are equivalent). Instead, in the case of the Pareto approach, this is literally what we are doing when writing Eq. \eqref{measureball1}, as we are remapping the radius of the ball in a monotonic fashion though the inverse of the $g$-functions.
\subsection{Relationship between the Gnedenko and Pareto approaches}
The relation between GEV and GPD parameters have been already discussed in literature in case of i.i.d variables  \citep{katz2005statistics,coles2001introduction,ding2008newly,malevergne2006power}. \citet{coles2001introduction} and \citet{katz2005statistics} have proven that the cdf of the GEV defined as $F_{GEV}(z; \mu, \alpha, \kappa)$ can be asymptotically written as that of GPD under a high enough threshold as follows:
\begin{eqnarray}
F_{GEV}(z; \mu, \alpha, \kappa) &\sim F_{GPD}(z;T,\sigma,\xi)=\nonumber \\ &= 1 - \left[ 1- \xi\left( \frac{z-T}{\sigma} \right) \right]^{1/\xi}
\end{eqnarray}
where $\kappa=\xi$, $\sigma=\alpha +\xi(T-\mu)$, and $T=\mu + \frac{\sigma}{\xi}(\lambda^{-\xi} -1)$, with $\ln(\alpha)=\ln(\sigma) + \xi\ln(\lambda)$.
In the present case, we have to compare Eqs. \eqref{g1res}-\eqref{g3res} for GEV with Eqs. \eqref{gpd1res}-\eqref{gpd3res} for GPD, keeping in mind that the GEV results hold only under the mixing conditions discussed before. While it is immediate to check that $\kappa=\xi$, the other relationships are valid in the limit of large $n$, as expected.
\section{\label{sec:numeric}Numerical Investigation}
The standard map \citep{chirikov} is an area-preserving chaotic map defined on the bidimensional torus, and it is one of the most widely-studied examples of dynamical chaos in physics.
The corresponding mechanical system is usually called a kicked rotator. It is defined as:
\begin{equation}
\begin{cases}
y_{t+1}= y_t - \frac{K}{2\pi}\sin(2 \pi x_t) & \mod 1  \\
x_{t+1}= x_t + y_t+1 & \mod 1  \\
\end{cases}
\label{stdmap}
\end{equation}
The dynamics of the map given in Eq. \eqref{stdmap} can be regular or chaotic. For $K<<1$ the motion follows quasi periodic orbits for all initial conditions, whereas if $K>>1$ the motion turns to be chaotic and irregular. An interesting behavior is achieved when $K \sim 1$: in this case we have coexistence of regular and chaotic motions depending on the chosen initial
conditions \citep{ott2002}.

\begin{figure*}
\includegraphics[width=0.95\textwidth]{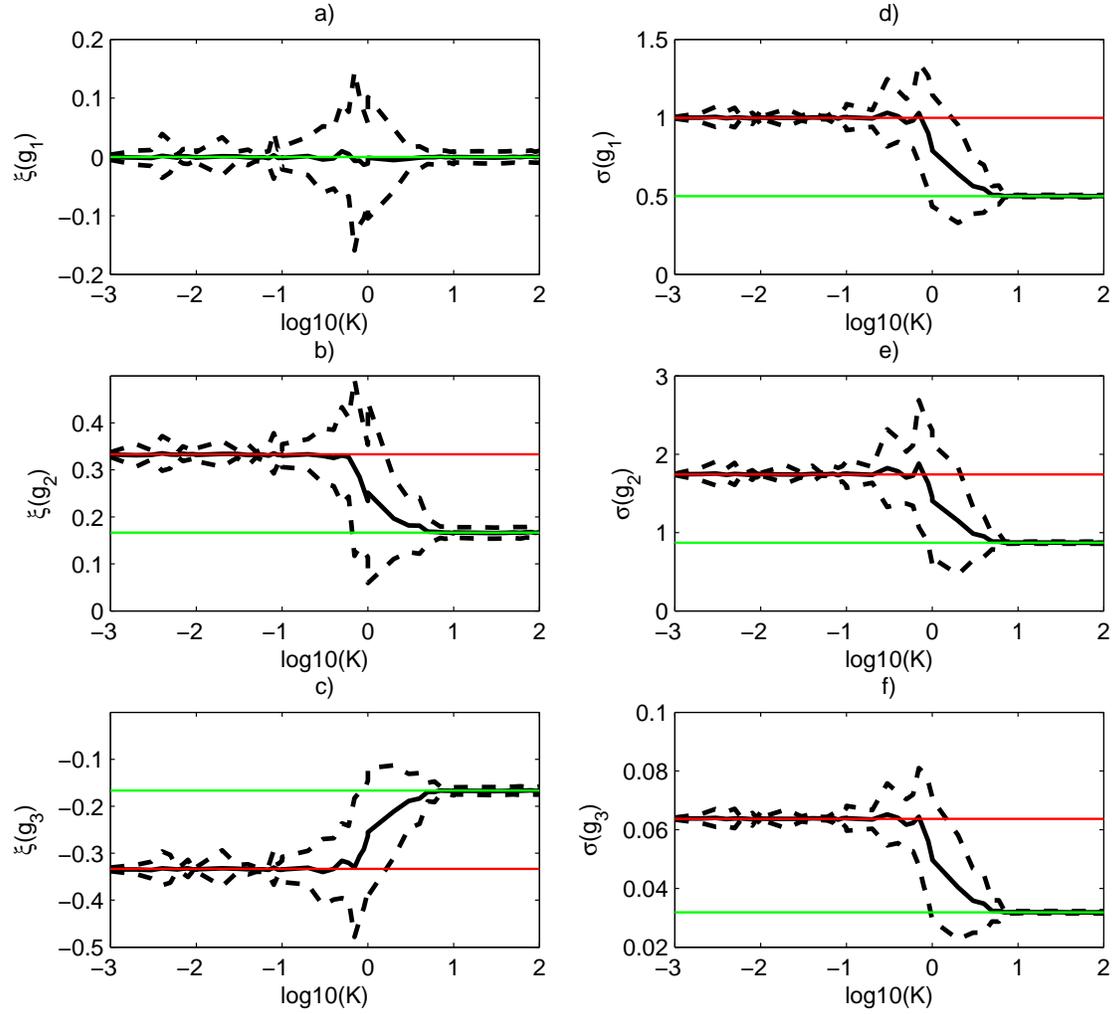}
\caption{GPD parameters for the observables $g_i$, $i=1,2,3$ computed over orbits of the standard map, for various values of the constant $K$. For each value of $K$, results refer to an esnsemble of 200 randomly chosen initial conditions $\zeta$. The notation $p(g_i)$ indicates the parameter $p$ computed using the extreme value statistics of the observable $g_i$. \textbf{a)} $\xi(g_1)$ VS $K$, \textbf{b)} $\xi(g_2)$ VS $K$, \textbf{c)} $\xi(g_3)$ VS $K$, \textbf{d)} $\sigma(g_1)$ VS $K$, \textbf{e)} $\sigma(g_2)$ VS $K$, \textbf{f)} $\sigma(g_3)$ VS $K$. Black solid lines: ensemble-average value. Black dotted lines: ensemble spread evaluated as one standard deviation of the ensemble. Green lines: theoretical values for regular orbits. Red lines: theoretical values for chaotic orbits.}
\label{fig1}
\end{figure*}

We perform for various values of $K$ ranging from $K=10^{-4}$ up to $K=10^2$ an ensemble of 200 simulations, each characterised by a different initial condition $\zeta$ randomly taken on the bidimensional torus, and we compute for each orbit the observables $g_i$, $i=1,2,3$. In each case, the map is iterated until obtaining a statistics consisting $10^4$ exceedances, where the threshold $T=7\cdot 10^{-3}$ and $\beta=3$. We have carefully checked that all the results are indeed robust with respect to the choice of the threshold and of the value of $\beta$. For each orbit, we fit the statistics of the $10^4$ exceedances values of the observables to a GPD distribution, using a MLE estimation \citep{castillo1997fitting} implemented in the MATLAB$^{\copyright}$ function \textit{gpdfit} \citep{matlab}. The results are shown in Fig. \ref{fig1} for the inferred values of $\xi$ and $\sigma$ and should be compared with Eqs. \eqref{gpd1res}-\eqref{gpd3res}. When $K\ll 1$, we obtain that the estimates of $\xi$ and $\sigma$ are compatible with a dimension $D=1$ for all the initial conditions: we have that the ensemble spread is negligible. Similarly, for $K\gg 1$, the estimates for $\xi$ and $\sigma$ agree remarkably well with having a local dimension $D=2$ for all the initial conditions. In the transition regime, which occurs for $K\simeq 1$, the ensemble spread is much higher, because the scaling properties of the measure is different among the various initial conditions. As expected, the ensemble averages of the parameters change monotonically from the value pertaining to the regular regime to that pertaining to the chaotic regime with increasing values of $K$. Basically, this measures the fact that the so-called regular islands shrink with $K$. Note that in the case of the observable $g_1$, the estimate of the  $\xi$ is robust in all regimes, even if, as expected, in the transition between low and high values of $K$ the ensemble spread is larger.  These results can also be compared with the analysis presented in \citet{faranda2011generalized}, where we used the Gnedenko approach. In that case, the values obtained in the regular regions were inconsistent with the GEV findings, the very reason being that the dynamics was indeed not mixing. Here, it is clear that the statistics can be computed in all cases, and we have a powerful method for discriminating regular from chaotic behaviors through the analysis of the inferred local dimension.
\section{Conclusions}
The growing attention of the scientific community in understanding the behavior of extreme values have led, in the recent past, to the development of an extreme value theory for dynamical systems. In this framework, it has been shown that the statistics of extreme value can be linked to the statistics of return in a neighborhood of a certain initial conditions by choosing special observables that depend on the distance between the iterated trajectory and its starting point. Until now, rigorous results have been obtained assuming the existence of an invariant measure for the dynamical systems and the fulfillment of independence requirement on the series of maxima achieved by imposing $D'$ and $D_2$ mixing conditions, or, alternatively, assuming an exponential hitting time statistics \citep{freitas2008,freitas,freitas2010extremal,gupta2009extreme}. The parameters of the GEV distribution obtained choosing as observables the function $g_i$, $i=1,2,3$ defined above depend on the local dimension of the attractor $D$ and numerical algorithms to perform statistical inference can be set up for mixing systems having both absolutely continuous and singular invariant measures \citep{faranda2011numerical,faranda2011extreme,vitolo2009robust,holland2011extreme}. Instead, when considering systems with regular dynamics, the statistics of the block maxima of any observable does not converge to the GEV family \citep{balakrishnan,nicolis}.

Taking a complementary point of view, in this paper we have studied the statistics of exceedances for the same class of observables and derived the limiting distributions assuming only the existence of an invariant measure and the possibility to define a local dimension $D$ around the point $\zeta$ of interest. To prove that the limiting distribution is a GPD we did not use any further conditions. In particular no assumptions on the mixing nature of the maxima sequence have been made. This means that a GPD limiting distribution holds for the statistics of exceedance of every kind of dynamical systems and it depends only on the threshold value and on the local dimension once we choose the observables $g_i$, $i=1,2,3$. Other functions can converge to the limiting behaviour of the GPD family if they asymptotically behave like the $g_i$'s (compare the discussion in \citep{freitas}). Nonetheless, this requires, analytically, to perform separately the limit for the threshold $T$ going to $g_{max}$ and that for the radius of the ball going to zero. In practical terms, this requires, potentially, much stricter selection criteria for the exceedances when finite time series are considered.

We note that, as the parameter $\xi$ is inversely proportional to $D$, one can expect that each time we analyse systems of intermediate or high dimensionality, the distributions for $g_2$ and $g_3$ observables will be virtually indistinguishable from what obtained considering the $g_1$ observable: the $\xi=0$ is in this sense an \textit{attracting} member of the GPD family. This may also explain, at least qualitatively, why the Gumbel ($k=0$ for the GEV family) distribution is so efficient in describing the extremes of a large variety of natural phenomena \citep{ghil2010extreme}.

The universality of this approach allows to resolve the debate on whether there exists or not a general way to obtain information about extreme values for quasi-periodic motions raised in \citet{balakrishnan} and \citet{nicolis}. This is due to the fact that considering  several of the largest order statistics instead of just the largest one we can study orbits where numerous exceedance are observed in a given block, as it happens for systems with periodic or quasi-periodic behaviors. As an example, one may consider a system with multiple commensurable frequencies: choosing a block length larger or equal to the smallest common period, we select always the same value for any considered observable. On the other hand for mixing maps we find, as expected, an asymptotic equivalence of the results obtained via the Gnedenko and via the Pareto approach.

The Pareto approach provides a  way to reconstruct local  properties of invariant measures: once a threshold is chosen and  a suitable exceedance statistics is recorded, we can compute the local dimension for different initial conditions taken on the attractor. This is true also in the opposite direction: if the knowledge of the exact value for the local dimension is available, once we chose a small enough radius (threshold), it is possible to compute a priori the properties of extremes without doing any further computations. In fact, the expression for the parameters Eqs. \eqref{gpd1res}-\eqref{gpd3res} do not contain any dependence on the properties of the dynamics except the local dimension.

Besides the analytical results, we have proved that Pareto approach is easily accessible for numerical investigations. The algorithm used to perform numerical simulations is versatile and computationally accessible: unlike the GEV algorithm that requires a very high number of iterations to obtain unbiased statistics, using the Pareto approach we can fix a priori a value for the threshold and the number of maxima necessary to construct the statistics. With the simulations carried out on the standard maps, we obtain meaningful results with a much smaller statistics with respect to what observed when considering the Gnedenko approach.

We hope that the present contribution may provide a tool that is not only useful for the analysis of extreme events itself, but also for characterising the dynamical structure of attractors by giving a robust way to compute the local dimensions, with the new possibility of embracing also the case of quasi-periodic motions.

Future investigations will include the systematic study of the impact on the extreme value statistics of adding stochastic noise to regular and chaotic deterministic dynamical systems, and the use of the Ruelle response theory \citep{ruelle98,ruelle2009} to study the modulation of the statistics of extremes due to changes in the internal or external parameters of the system, especially in view of potentially relevant applications in geophysics such as in the case of climate studies \citep{abramov2007,lucarini2011}.
\begin{acknowledgments}
The authors acknowledge various useful exchanges with R. Blender and K. Fraedrich, and the financial support of the EU-ERC project NAMASTE-Thermodynamics of the Climate System.
\end{acknowledgments}
\bibliography{indicatori}
\end{document}